# Study of domain switching using piezoresponse force microscopy in $Ca_{0.4}Sr_{0.6}Bi_4Ti_4O_{15}$ thin film for electromechanical applications

Supratim Mitra[1*], Satakshi Gupta[2], Aneesh M. Joseph[3], Umesh Kumar Dwivedi[2]

[1]Department of Physics, BanasthaliVidyapith, Banasthali-304022, Rajasthan, India

[2]Amity School of Applied Sciences, Amity University, Jaipur-303002, Rajasthan, India

[3]IIT Bombay Nanofabrication Facility, Indian Institute of Technology Bombay, Mumbai-400076, India

*Corresponding Author: msupratim@banasthali.ac.in

## ABSTRACT

An attempt has been made to synthesize $(Ca_{0.4}Sr_{0.6})Bi_4Ti_4O_{15}$(CSBT) thin film using pulsed laser deposition (PLD) and successfully optimized the deposition conditions. A good quality film with the desired phase is obtained at a substrate temperature of 650 °C based on phase, composition and morphology studies. Highly *c*-axis oriented films with average thickness 330-400 nm and an average grain size 40-65 nm has been found. Piezoresponse force microscopy (PFM) showed a complete domain reversal using switching spectroscopy. A comparatively high effective $d_{33}^*$ value as ~120 pm/V has been achived. The results suggest that CSBT has a great potential in electromechanical applications.

## 1.Introduction

Bismuth layer-structured ferroelectrics (BLSFs) have drawn much interest of the researchers due to their application in thin film memory devices, high-temperature electromechanical devices owing to their high Curie temperature ($T_C$), good ferroelectric properties, good fatigue resistance andafter all their lead-free nature [1]. Sensors, actuators, and transducers operating at elevated temperatures (~400 °C) are highly desired in industries such as energy, engines, oil and gas, and chemicals. In these applications, the operating temperature should not exceed half of the Curie-temperature ($T_C$) of piezoelectric materials. Piezoelectric materials based on BLSFs are the potential candidate to operate at high-temperature and provide intended functionality. BLSFs belong to Aurivillius phase with the general chemical formula $(Bi_2O_2)^{2+}$ $(A_{m-1}B_mO_{3m+1})^{2-}$ where A is one or more mono-, di-, or trivalent ions ($Sr^{2+}$, $Ba^{2+}$, $Ca^{2+}$, $Bi^{3+}$ etc. or combination), B is tetra-, penta- or hexavalent ions ($Ti^{4+}$, $Ta^{5+}$, $Nb^{5+}$, etc.) and $m$ ( = 1, 2, 3, 4, etc.) is the number of the $BO_6$octahedra regularly interleaved by $(Bi_2O_2)^{2+}$ layer along $c$-axis[2]. The piezoelectric propertiesinthese BLSF compounds are greatly influenced by the number of layers and the elements present in the A and B site [1]. Thus, the properties of these compounds can be enhanced by incorporating different cations at site A or B or both [3]. However, it has been reported that the Curie temperature of BLSFs decreases when the size of the cations at A-site increases, while Curie temperature increases and the electrical properties are augmented significantly due to the replacement of A- site cation by a smaller size ion [4]. In addition, most of the BLSFs possess a relatively high electrical conductivity which makes the poling process difficult.

Among the various BLSFs are studied, the compounds, $M Bi_4Ti_4O_{15}$ with $m = 4$ and $M =$ Sr, Ca, Ba have been investigated extensively due to their better ferroelectric properties [2, 4-10]. CaBi$_4$Ti$_4$O$_{15}$ (CBT) and SrBi$_4$Ti$_4$O$_{15}$ (SBT) have been studied mostly due to their strong anisotropic character, high remnant polarization and high $T_C$. CBT isrecognizedfor itscomparativelyhigh $T_C$ (790 $^o$C), low electrical conductivity, and low dielectric loss, however, with a low piezoelectric coefficient (8-9 pC/N)[8, 11, 12]. Further,whenCa$^{2+}$ is partially replaced by Sr$^{2+}$, the composition (Ca$_{0.4}$Sr$_{0.6}$)Bi$_4$T$_4$O$_{15}$ (CSBT) showed enhanced electrical properties ($d_{33}$ ~ 15 pC/N)[8]. Therefore, CSBT is expected to be a potential candidate for electromechanical applications even at moderately high-temperature environment (~ 350 $^o$C).

Synthesis of BLSFs in thin film form is of practical importance in various situations and particularly for the design of piezoelectric devices. These thin films are of interest because of their added advantage of strong piezoelectric effect which allows electromechanical sensing and actuation. However, the successful growth of such a film is very challenging. In addition, investigation of the domain structure and polarization switching are very fundamental for piezo/ferroelectric thin film and extremely necessary for electromechanical device applications [13, 14]. Electromechanical properties in these films are also influenced by composition and orientation of grain. Among the various techniques used to evaluate the electromechanical functionalities in these ferroelectric and piezoelectric thin film, characterization using piezoresponse force microscopy (PFM) has been proven a useful tool to explore the static and dynamic electromechanical performances in terms of domain morphology, structure, growth, and polarization switching in nanoscale[15-19]. Attempts have been made to study variousSBT, CBT thinfilmsmainly focusing on the dielectric, ferroelectric properties and fatigue resistance[20-26] for mainly memory application, however, design and understanding of piezoelectric properties of (Ca$_{0.4}$Sr$_{0.6}$)Bi$_4$T$_4$O$_{15}$ (CSBT) thin film for potential high-temperature electromechanical applicationsarehardly reported.

Therefore, in this letter, we have attempted to synthesis $(Ca_{0.4}Sr_{0.6})Bi_4T_4O_{15}$ thin film using pulsed laser deposition (PLD), followed by an effort on optimization of deposition conditions by phase, composition and morphology studies. The quality of the film and piezoelectric properties were characterized using atomic force microscopy (AFM) and piezoresponse force microscopy (PFM) to realize domain switching as potential applications in electromechanical devices.

## 2. Experimental

The thin film of $Ca_xSr_{1-x}Bi_4T_4O_{15}$ ($x = 0.4$) (CSBT) was deposited in Pt/CSBT/Pt (111) configuration on anRCA cleaned platinized Si (100) substrate (Pt(111)/TiO$_2$/SiO$_2$/Si)) using PLD[#]. The PLD was equipped with a Coherent compex Pro 201, 248 nm KrF excimer laser with frequency = 5 Hz, laser energy = 2 J/cm$^2$, pulse width = 30 nsec. The optimized target to substrate distance = 4 cm, and spot size = 3×1.5 mm$^2$ were kept fixed during all the depositions. The films were deposited at a different substrate temperature (650, 700, 750 $^o$C) and oxygen pressure in the chamber (0.2, 0.3, 0.5 mbar). After deposition, the films were annealed at the corresponding deposition temperature for 10 min and then slowly cooled down to room temperature. The cooling rate was kept slow (cooling rate was 2°C/min for first 200°C and then about 3°C/min for the rest) so that the film does not develop cracks due to the thermal shock originated at the interface of CSBT film and the substrate because of rapid cooling. At the time of annealing after the deposition, the oxygen pressure was kept at 0.5 atm to reduce the oxygen deficiencies which adversely affect the electrical properties of the film. The thickness of the as-deposited film was measured with help of Profilometer (Bruker Dektak XT). Finally, the top electrode Pt was sputter deposited using a shadow mask. The thickness of Pt layer was determined with the help of Ellipsometer (Sentech Instruments SE 800). The phase of the films was identified using High-resolution X-Ray Diffractometer (HR-XRD, RigakuSmartlab 3KW)with Cu $K\alpha$ ($\lambda$ = 1.5406 Å). Surface morphology and microstructure was observed using SEM (Raith150 TWO tool) and AFM in tapping mode (Asylum Research, Oxford Instruments). The piezoelectric response and domain switching effects were recorded in PFM mode (Asylum Research, Oxford Instruments). For this, a conductive cantilever of spring constant 2 N/m and a Ti-Ir coated tip of radius 28 nm was used to visualize the domain.

## 3. Results and discussion

**Figure 1(a)** depicts the XRD patterns in a θ-2θ scan of the CSBT thin film deposited at various deposition conditions (substrate temperature = 650, 700, 750 °C, pressure = 0.2, 0.3, 0.5 mbar) on a platinized Si substrate (Pt(*111*)/TiO$_2$/SiO$_2$/Si(*100*)). The XRD pattern of CSBT target (bulk)with an orthorhombic symmetry (*A 2$_1$a m*) is also recorded and shown in **figure 1(b)**for comparison. As observed in **figure 1(a)**, a complete phaseofCSBTfilmhas formed at the substrate temperature of 650 °Candall the detected peaks are found to be in good agreement with the bulk sample. The identified strong peaks (*00l*) (*l* = 6, 8, 10, 20), (*119*)suggest that the film has the higher amount of *c*-axis oriented grains compared to the bulk. The *c*-axis ratio of the film is calculated using, *α = I(0 0 10)/{I(0 0 10) + I(2 2 0)}*and found to be 0.78, 0.80, 0.91 for the films deposited at the temperature 650 °C and pressure 0.5, 0.3, 0.2 mbar respectively[27]. In general, the higher deposition temperature is desired as to avoidthe appearance of intermediate phases such as flurite and pyrochlore[28]. However, in CSBT thin film an incomplete phase formation was observed for the films that were deposited at temperature 700 and 750 °C suggesting a specific energy window required for phase formation and crystalline orientation.Theincomplete phase formation in the films deposited at 700 and 750 °C suggest thattheactivation energy required for grain growth of (*119*) is low compared to grain growth of (*00l*)-oriented film, which results in the growth of non (*00l*)-oriented film[20]. On the other hand, the influence of substrate on crystal structure could not be discarded completely and expected to destabilize the columnar *<111>* textured Pt electrode at temperature 700 and 750 °C, which might result in incomplete phase formation[29].The composition of CSBT film deposited at 650 °C was confirmed by the energy dispersive spectroscopy (EDS), whichshows the presence of all elements (Ca, Sr, Bi, Ti, O) in desired stoichiometry and their homogeneous distribution as shown in **figure 1(c)** as color elemental mapping.

    **Figure 2** shows the surface morphology of the CSBTthin films. SEM micrographs (shown at *right* for few representative deposition conditions) reveal that the film has a uniform, crack-free and dense grain. However, the average grain size ranges in the nanometer scale differ owing to the different temperature and pressure conditions. The average grain size has been calculated using the linear intercept method and the calculated average grain size at various substrate temperatures is shown in **figure 2** (*middle*). The average grain size of the films deposited at 650 °C is found to be the lowest ranges from 41 to 65 nm from 0.5 to 0.2 bar pressure, which further increases with increasing temperature at a constant pressure but decreases with increasing pressure at a constant substrate temperature. The 3-dimensional AFM topography images also confirm the films are uniformly deposited as shown in **figure 2** (*left*). The root mean square (RMS) value of roughness of the CSBT films is also calculated. The minimum RMS roughness value is found to be 4.23 to 7.13 nm for 0.5 to 0.2 mbar pressure for the films deposited at a substrate temperature of 650 °C. As the deposition temperature reaches 750 °C, the RMS roughness also increases to 19.88 nm. The measured thickness of the CSBT film deposited at different deposition conditions (substrate temperature and pressure) has been found in the range of 334-400 nm for all the samples[#]. Thus, based on the above analysis, films

only deposited at 650 °C and 0.2, 0.3, 0.5 mbar are found to be the best quality films with desired phase and morphology and therefore further characterized to investigate piezoelectric properties using PFM.

**Figure 3 (a-c)** depicts the simultaneously obtained surface topography, PFM phase and amplitude image of the as-grown CSBT film deposited at a substrate temperature of 650°C respectively. It is clear that the film has a uniform surface and defined grain boundary. Moreover, by merging surface topology and PFM phase image, it can be seen that most of the domain boundaries coincide with grain boundaries. In order to understand the ferroelectric domain switching on CSBT films, the writing was performed on a $2 \times 2$ μm$^2$ region, which was first negatively poled (-10V) and then positively poled (+10V) through the tip and read with a $6 \times 6$ μm$^2$ region as shown in **Figure 3(d-f)**. The negatively poled region shows a darker contrast and the positively poled region shows a brighter contrast as compared to the unpoled region. This color contrast manifests effective ferroelectric domain switching.

To further understand the domain switching process, PFM hysteresis loops were obtained for amplitude and phase with respect to applied voltage as shown in **figure 4(a-b)**. It is clear from **figure 4(a-b)** that a phase reversal of almost 180° is achieved in all the samples. This confirms the complete polarization reversal or domain switching. In addition, the amplitude-voltage loops also substantiate the domain reversal process. The effective piezoelectric coefficient ($d_{33}^*$) hysteresis loop wasconstructed from amplitude-voltage butterfly loop using the method described in Ref. [30]and shown in **figure 4(c)**. Both amplitude-bias voltage and $d_{33}^*$-bias voltage loops are slightly asymmetric in nature and shifted from the origin which could be attributed to the internal bias field [15, 31]. A maximum $d_{33}^*$ value of ~120 pm/V was obtained after tracing several loops and negligible variation found for other deposition conditions (0.2, 0.3, 0.5 mbar). The value of $d_{33}^*$ (~120 pm/V) obtained in this work is found to be higher as compared to the recently reported highest value (~60 pm/V) for CBT thin film[32], and much larger value (~15 pC/N) compared to the bulk CSBT sample [8]. However, the value reported by Fu *et al.* for sol-gel derived CBT film on Pt foil showed extremely large $d_{33}^*$ value as 180 pm/V [33].

## 4. Conclusions

In conclusion, an attempt has been made to synthesize $(Ca_{0.4}Sr_{0.6})Bi_4T_4O_{15}$(CSBT) thin film using pulsed laser deposition (PLD) and successfully optimized the deposition conditions. Phase, compositional, and morphology study enabled to choose substrate temperature as low as 650 °C. Piezoelectric properties were characterized using atomic force microscopy (AFM) and piezoresponse force microscopy (PFM). Observation of smaller grain size (40-65 nm), a complete domain reversal and comparatively high effective $d_{33}^*$ value as ~120 pm/V suggests CSBT has a potential for electromechanical applications.

#Detail can be found in **Supplementary Information**.

**Availability of data and material**

The data will be available without restriction from corresponding author

**Competing interests**

The Authors declare that they do not have competing interests

**Funding** The work is supported by Department of Science and Technology-Science and Engineering Research Board, India under Early Career Research Award scheme (Grant No. ECR/2016/000794/ES)

**Authors' contributions**

SG is a Master student carried out the experimental work with AMJ at INUP-IIT Bombay, India. SM is the group leader, has done the data analysis and jointly written and reviewd the manuscript with UKD.

**Acknowledgements**

Author SM would like to gratefully acknowledge INUP-IIT Bombay for carrying out the entire experimental work.

**Authors' information**

Not Applicable

**References**

[1] H. Nagata, T. Takenaka, Ferroelectrics 273(1) (2002) 359-364.

[2] G. Li, L. Zheng, Q. Yin, B. Jiang, W. Cao, Journal of Applied Physics 98(6) (2005) 064108.

[3] M.B. Suresh, E.V. Ramana, S.N. Babu, S.V. Suryanarayana, Ferroelectrics 332(1) (2006) 57-63.

[4] M.S. Islam, J. Kano, Q.R. Yin, S. Kojima, Journal of Electroceramics 28(2) (2012) 89-94.

[5] G.M. V., T. Daniel, E.J. A., Journal of the American Ceramic Society 82(9) (1999) 2368-2372.

[6] Y. Haixue, L. Chengen, Z. Jiaguang, Z. Weimin, H. Lianxin, S. Yuxin, Y. Youhua, Japanese Journal of Applied Physics 40(11R) (2001) 6501.

[7] M.D. Maeder, D. Damjanovic, C. Voisard, N. Setter, Journal of Materials Research 17(6) (2011) 1376-1384.

[8] Z. Liaoying, L. Guorong, Z. Wangzhong, Y. Qinrui, Japanese Journal of Applied Physics 41(12B) (2002) L1471.

[9] S.-T. Zhang, B. Sun, B. Yang, Y.-F. Chen, Z.-G. Liu, N.-B. Ming, Materials Letters 47(6) (2001) 334-338.

[10] R. Nie, Q. Chen, H. Liu, J. Xing, J. Zhu, D. Xiao, Journal of Materials Science 51(11) (2016) 5104-5112.

[11] L. Korzunova, Ferroelectrics 134(1) (1992) 175-180.

[12] A. Garg, Z.H. Barber, M. Dawber, J.F. Scott, A. Snedden, P. Lightfoot, Applied Physics Letters 83(12) (2003) 2414-2416.

[13] V. Shur, E. Rumyantsev, S. Makarov, Journal of Applied Physics 84(1) (1998) 445-451.

[14] N. Setter, D. Damjanovic, L. Eng, G. Fox, S. Gevorgian, S. Hong, A. Kingon, H. Kohlstedt, N.Y. Park, G.B. Stephenson, I. Stolitchnov, A.K. Taganstev, D.V. Taylor, T. Yamada, S. Streiffer, Journal of Applied Physics 100(5) (2006) 051606.

[15] H. Huang, X.L. Zhong, S.H. Xie, Y. Zhang, J.B. Wang, Y.C. Zhou, Journal of Applied Physics 110(5) (2011) 054105.

[16] D.J. Kim, J.Y. Jo, T.H. Kim, S.M. Yang, B. Chen, Y.S. Kim, T.W. Noh, Applied Physics Letters 91(13) (2007) 132903.

[17] A. Wu, P.M. Vilarinho, D. Wu, A. Gruverman, Applied Physics Letters 93(26) (2008) 262906.

[18] A. Gruverman, D. Wu, J.F. Scott, Physical Review Letters 100(9) (2008) 097601.

[19] L. Keeney, P.F. Zhang, C. Groh, M.E. Pemble, R.W. Whatmore, Journal of Applied Physics 108(4) (2010) 042004.

[20] J. Yan, G. Hu, Z. Liu, S. Fan, Y. Zhou, C. Yang, W. Wu, Journal of Applied Physics 103(5) (2008) 056109.

[21] H. Sun, J. Zhu, H. Fang, X.-b. Chen, Journal of Applied Physics 100(7) (2006) 074102.

[22] D.S. Sohn, W.X. Xianyu, W.I. Lee, I. Lee, I. Chung, Applied Physics Letters 79(22) (2001) 3672-3674.

[23] A. Gruverman, A. Pignolet, K.M. Satyalakshmi, M. Alexe, N.D. Zakharov, D. Hesse, Applied Physics Letters 76(1) (1999) 106-108.

[24] J.S. Liu, S.R. Zhang, L.S. Dai, Y. Yuan, Journal of Applied Physics 97(10) (2005) 104102.

[25] S.-T. Zhang, B. Yang, Y.-F. Chen, Z.-G. Liu, X.-B. Yin, Y. Wang, M. Wang, N.-B. Ming, Journal of Applied Physics 91(5) (2002) 3160-3164.

[26] K. Kato, K. Tanaka, K. Suzuki, T. Kimura, K. Nishizawa, T. Miki, Applied Physics Letters 86(11) (2005) 112901.

[27] K. Iijima, Y. Tomita, R. Takayama, I. Ueda, Journal of Applied Physics 60(1) (1986) 361-367.

[28] T.J. Boyle, C.D. Buchheit, M.A. Rodriguez, H.N. Al-Shareef, B.A. Hernandez, B. Scott, J.W. Ziller, Journal of Materials Research 11(9) (2011) 2274-2281.

[29] H. Wang, L.W. Fu, S.X. Shang, X.L. Wang, M.H. Jiang, Journal of Physics D: Applied Physics 27(2) (1994) 393.

[30] Y.C. Yang, C. Song, X.H. Wang, F. Zeng, F. Pan, Applied Physics Letters 92(1) (2008) 012907.

[31] S. Mitra, T. Karthik, J. Kolte, R. Ade, N. Venkataramani, A.R. Kulkarni, Scripta Materialia 149 (2018) 134-138.

[32] A.Z. Simões, M.A. Ramírez, A. Ries, J.A. Varela, E. Longo, R. Ramesh, Applied Physics Letters 88(7) (2006) 072916.

[33] D. Fu, K. Suzuki, K. Kato, Applied Physics Letters 85(16) (2004) 3519-3521.

**Figure Captions**

**FIGURE 1:** XRD pattern in a θ-2θ scan of (a) CSBT thin film samples prepared at different temperature and pressure conditions, (b) bulk CSBT sample, (c) EDS elemental mapping of CSBT film.

**FIGURE 2:** Variation of average grain size and roughness with a substrate temperature (650, 700, 750°C) for CSBT film deposited on Pt/Si substrate at 0.2, 0.3, 0.5 mbar. Representative (*left*) 3-dimensional AFM images, (*right*) SEM micrographs for CSBT film deposited at different temperature and pressure conditions.

**FIGURE 3:** (a) Surface topography, (b) PFM phase image, (c) PFM amplitude for the CSBT film grown at 650°C, 0.5 mbar. Ferroelectric domain switching for CSBT thin film grown on Pt/Si substrate at 650°C and 0.3 mbar. (d) Typical surface topography, PFM phase image after poling with D.C. bias of (e) -10V and (f) +10 V.

**FIGURE 4:** PFM (a) amplitude-voltage curve, (b) phase-voltage hysteresis curve (c) effective piezoelectric coefficient ($d_{33}^*$)-voltage hysteresis curvefor CSBT film deposited on Pt/Si substrate at 650°C and 0.3 mbar.

Figure 1

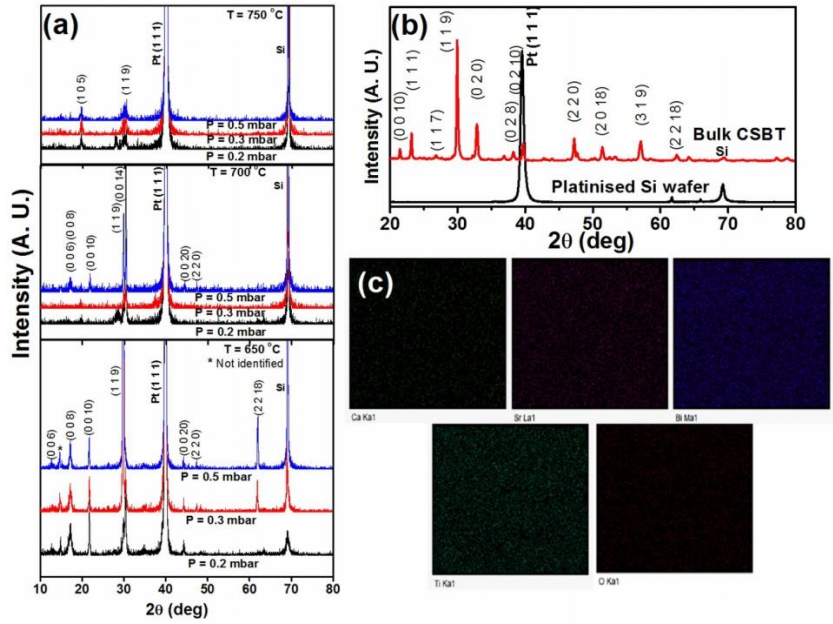

Figure 2

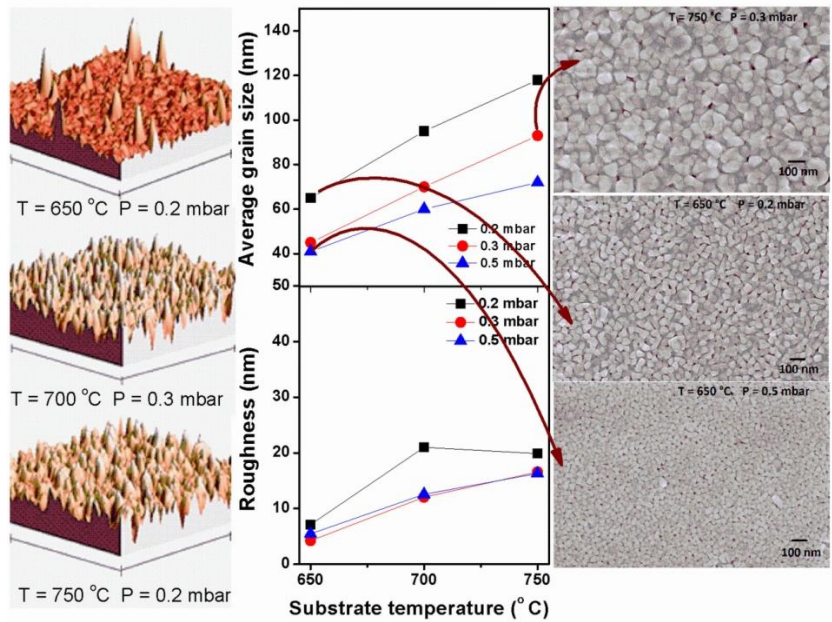

Figure 3

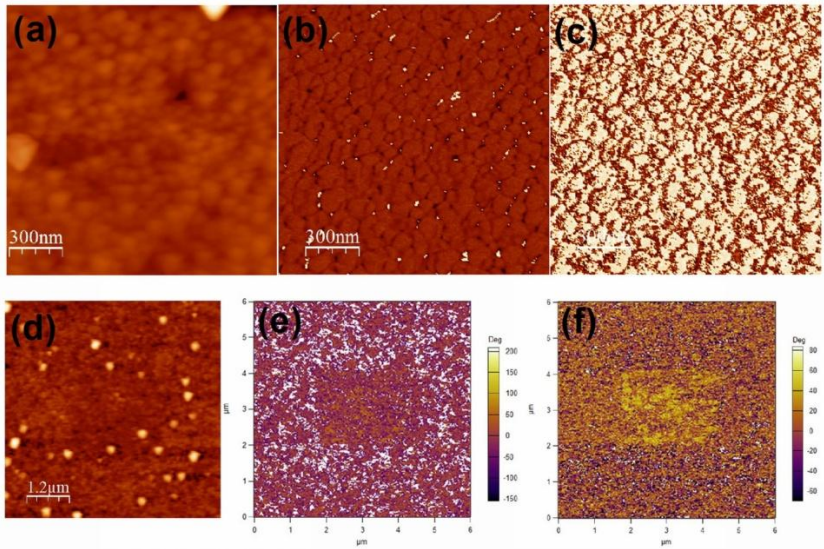

Figure 4

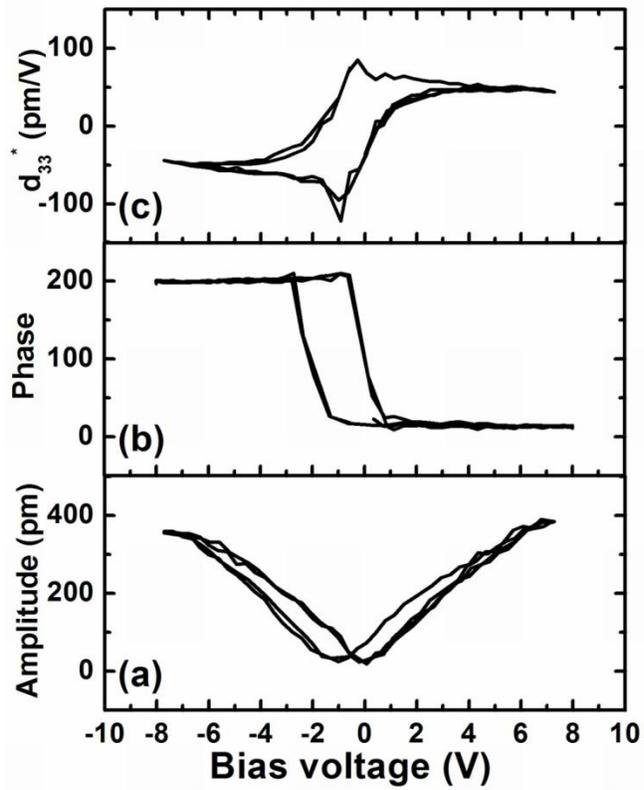